\documentclass[aps,twocolumn,pra,superscriptaddress,showpacs,tightenlines]{revtex4}
\usepackage{epsfig,graphicx,times}
\usepackage{amstext}
\usepackage{amsmath}
\usepackage{amssymb}
\usepackage{graphicx}
\usepackage{latexsym}
\usepackage{bm}
\usepackage[colorlinks,citecolor=blue, linkcolor=blue,hyperindex,CJKbookmarks,dvipdfm]{hyperref}

\begin{document}

\title{Phonon blockade in a nanomechanical resonator resonantly coupled to a
qubit}
\author{Xun-Wei Xu}
\email{davidxu0816@163.com}
\affiliation{Department of Applied Physics, East China Jiaotong University, Nanchang,
330013, China}
\author{Ai-Xi Chen}
\email{aixichen@ecjtu.edu.cn}
\affiliation{Department of Applied Physics, East China Jiaotong University, Nanchang,
330013, China}
\author{Yu-xi Liu}
\affiliation{Institute of Microelectronics, Tsinghua University, Beijing 100084, China}
\affiliation{Tsinghua National Laboratory for Information Science and Technology
(TNList), Beijing 100084, China}
\date{\today }

\begin{abstract}
We study phonon statistics in a nanomechanical resonator (NAMR) which is
resonantly coupled to a qubit. We find that there are two different
mechanisms for phonon blockade in such a resonantly coupled NAMR-qubit
system. One is due to the strong anharmonicity of the NAMR-qubit system with
large coupling strength; the other one is due to the destructive
interference between different paths for two-phonon excitation in the NAMR-qubit system with a
moderate coupling strength.
We find that the phonon blockade is fragile towards thermal mode occupations and
can only be observed for NAMR being at ultracold effective temperature.
In order to enlarge the mean phonon
number for strong phonon antibunching with a moderate NAMR-qubit coupling strength,
we assume that two external driving fields are applied to the NAMR and qubit, respectively. In this case, we find
that the phonon blockades under two mechanisms can appear at the same
frequency regime by optimizing the strength ratio and phase difference of the two external
driving fields.
\end{abstract}

\maketitle




\section{Introduction}

Quantum effects in a nanomechanical resonator (NAMR) can be explored when
the vibration energy of NAMR can beat the thermal energy and approach the
quantum limit. Recently, the experiment showed that the quantum limit has
been reached in a NAMR with a sufficiently high frequency at a very low
temperature~\cite{OConnellNat10}. This makes it possible for NAMR to be
applied to quantum information processing~\cite{SchwabPT05}.

If a NAMR approaches the quantum regime and the quanta of the mechanical
oscillation, referred to as phonons, can be generated one by one, then a
purely quantum phenomenon, phonon blockade, can be explored. In analogy to
the Coulomb blockade~\cite{KastnerPT93} and photon blockade~\cite%
{ImamogluPRL97}, the phonon blockade~\cite{YXLiuPRA10} is a phenomenon that
only one phonon can be excited in a nonlinear mechanical oscillator~\cite{BarzanjehPRA16} by
external driving fields. Phonon blockade has already been studied in a NAMR
coupled to a superconducting qubit in the dispersive regime~\cite%
{YXLiuPRA10,DidierPRB11,MiranowiczPRA16,XWangARX16}. It should be more
easily observed for larger nonlinear phonon interaction induced by the
qubit, which corresponds to a larger coupling strength and moderate detuning
between the NAMR and the qubit~\cite{YXLiuPRA10}.

In contrast to Refs.~\cite{YXLiuPRA10,DidierPRB11,MiranowiczPRA16,XWangARX16}, where the NAMR is dispersively coupled to the qubit, here we are going to study the phonon statistics in a NAMR
which is resonantly coupled to a qubit. In recent several works, a similar
system, in which a single two-level defect system is coupled to an
optomechanical system~\cite{RamosPRL13,HWangPRA15}, was proposed. Such a
system can be used to realize phonon blockade due to the strong
anharmonicity of the eigenstates corresponding to large NAMR-qubit coupling
strength~\cite{RamosPRL13} in analogy to cavity QED~\cite%
{BirnbaumNat05,FaraonNPy08,CLangPRL11,HoffmanPRL11}, and we call it as
conventional phonon blockade (CPNB). Different from the previous studies~%
\cite{RamosPRL13}, we here show that phonon blockade in the resonantly
coupled NAMR-qubit system can also be produced by the destructive
interference between different paths for two-phonon excitation with a
moderate NAMR-qubit coupling strength. We call the interference-based phonon blockade as unconventional phonon
blockade (UCPNB), which is similar to the unconventional photon
blockade in a weakly nonlinear system of photonic molecule~\cite%
{LiewPRL10,BambaPRA11,LemondePRA14,MajumdarPRL12,KomarPRA13,GeracePRA14,KyriienkoPRA14a,XuPRA14a,XuPRA14b,KyriienkoPRA14,ShenPRA15,ZhouPRA15}%
. Moreover, we find that the phonon blockade is fragile towards thermal mode occupations for the small mean phonon number and can only be observed for NAMR being at ultracold effective temperature.

To improve the robustness of the phonon blockade against the thermal
noise and to increase the number of single phonons generated in a given time,
we discuss how to enlarge the mean phonon number for strong
phonon antibunching with a moderate NAMR-qubit coupling. Recently, photon
blockade effect in a quantum dot-cavity system for the cavity and the
quantum dot driven by two external fields respectively was studied~\cite%
{TangSR15}, and an enhancement of photon blockade with large photon number
was achieved with some optimized parameters for the two driving fields. In
the spirit of the approach of Ref.~\cite{TangSR15}, we assume that two external driving fields are applied to the NAMR and qubit,
respectively, then a large mean phonon number for strong phonon antibunching is obtained with a moderate NAMR-qubit
coupling strength. The optimal detuning for CPNB is mainly determined by the
NAMR-qubit coupling strength, while the optimal detuning for UCPNB is related
to the strength ratio and phase difference of the two external driving fields. A large mean
phonon number for strong phonon antibunching with a moderate NAMR-qubit
coupling strength can be obtained by combining the effects of CPNB and UCPNB when
they appear at the same frequency regime by optimizing the strength ratio and phase difference of the two external driving fields.

The paper is organized as follows. In Sec.~II, we describe the theoretical
model of the resonant coupling between a NAMR and a qubit, and then we give
analytical results of the optimal conditions for observing phonon
blockade. In Sec.~III, we study the phonon blockade when the external
driving field is not applied to the qubit. In Sec.~IV, we consider how to
enlarge the mean phonon number for the phonon blockade by an external
driving field applied to the qubit. In Sec.~V, we study how to detect the phonon
blockade by an optical cavity which is optomechanically coupled to the
NAMR. Finally, the main results of our work are summarized in Sec.~VI.

\section{Theoretical model and analytical results}

\begin{figure}[tbp]
\includegraphics[bb=23 271 455 643, width=7 cm, clip]{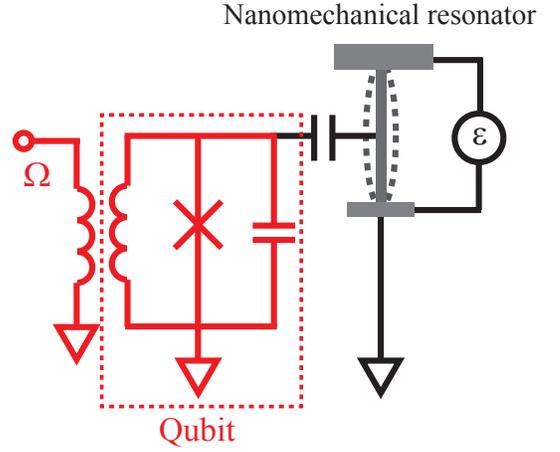}
\caption{(Color online) Schematic diagram that
a nanomechanical resonator (gray) is coupled to a phase qubit (red). The
nanomechanical resonator and phase qubit are driven by external fields with
strengths $\protect\varepsilon$ and $\Omega$, respectively.}
\label{fig1}
\end{figure}

Motivated by a recent experiment~\cite%
{OConnellNat10}, we study the phonon blockade in a quantum
system, in which a high-frequency NAMR is coupled to a phase qubit, as
schematically shown in Fig.~\ref{fig1}. The Hamiltonian of the whole system is given by ($\hbar =1$)%
\begin{eqnarray}
H_{\mathrm{mq}} &=&\omega _{0}\sigma _{+}\sigma _{-}+\omega _{m}b^{\dag
}b+J\left( \sigma _{+}b+b^{\dag }\sigma _{-}\right)  \notag \\
&&+\left( \Omega e^{-i\phi }e^{-i\omega _{q}t}\sigma _{+}+\varepsilon
b^{\dag }e^{-i\omega _{b}t}+\mathrm{H.c.}\right)
\end{eqnarray}%
 under the rotating wave approximation, where $\sigma _{+}$ and $\sigma _{-}$ are the raising and lowering
operators of the qubit with frequency $\omega _{0}$; $b$ and $b^{\dag }$ denote the
annihilation and creation operators of the NAMR with frequency $\omega _{m}$%
; $J$ is the NAMR-qubit coupling strength; $\Omega $ ($\varepsilon $)
describes the coupling strength between tha qubit (NAMR) and the external driving field
with frequency $\omega _{q}$ ($\omega _{b}$) and $\phi $ is the phase
difference between the two external driving fields. Hereafter we assume that $\Omega $ and $\varepsilon $ are real numbers and the frequencies of the two driving fields are the same and $ \omega _{q}=\omega _{b}=\omega _{d}$. In this paper, we focus on the phonon blockade for the resonant NAMR-qubit coupling, i.e., $\omega _{0}=\omega _{m}$. In the rotating reference frame with the
frequency $\omega _{d}$ of the driving fields, the Hamiltonian of the quantum system is shown as%
\begin{eqnarray}
H_{\mathrm{mq}}^{\prime } &=&\Delta \sigma _{+}\sigma _{-}+\Delta b^{\dag
}b+J\left( \sigma _{+}b+b^{\dag }\sigma _{-}\right)  \notag  \label{Eq2} \\
&&+\left( \Omega e^{-i\phi }\sigma _{+}+\varepsilon b^{\dag }+\mathrm{H.c.}%
\right)
\end{eqnarray}%
with detuning $\Delta \equiv \omega _{0}-\omega _{d}=\omega _{m}-\omega _{d}$.

The second-order correlation function of the phonons in the steady state
(i.e., $t\rightarrow +\infty $) is defined by
\begin{equation}
g_{b}^{\left( 2\right) }\left( \tau \right) \equiv\frac{\left\langle b^{\dag
}\left( t\right) b^{\dag }\left( t+\tau \right) b\left( t+\tau \right)
b\left( t\right) \right\rangle }{n_{b}^{2}},
\end{equation}%
where $n_{b}\equiv\left\langle b^{\dag }\left( t\right) b\left( t\right)
\right\rangle $ is the mean phonon number in the NAMR. The correlation
function of the phonons can be calculated by numerically solving the master
equation for the density matrix $\rho $ of the system~\cite{Carmichael}
\begin{eqnarray}
\frac{\partial \rho }{\partial t} &=&-i\left[ H_{\mathrm{mq}}^{\prime },\rho %
\right]   \notag  \label{Eq6} \\
&&+\gamma \left( n_{m,\mathrm{th}}+1\right) L[b]\rho +\gamma n_{m,\mathrm{th}%
}L[b^{\dag }]\rho  \\
&&+\kappa \left( n_{q,\mathrm{th}}+1\right) L[\sigma _{\_}]\rho +\kappa n_{q,%
\mathrm{th}}L[\sigma _{+}]\rho ,  \notag
\end{eqnarray}%
where $L[o]\rho =o\rho o^{\dag }-\left( o^{\dag }o\rho +\rho o^{\dag
}o\right) /2$ denotes a Lindbland term for an operator $o$, $\kappa $ is
damping rate of the qubit and $\gamma $ is damping rate of the NAMR; $n_{q,%
\mathrm{th}}$ and $n_{m,\mathrm{th}}$ are the mean numbers of the thermal
photon (phonon), given by the Bose-Einstein statistics $n_{q,\mathrm{th}%
}=[\exp (\hbar \omega _{0}/k_{B}T)-1]^{-1}$, $n_{m,\mathrm{th}}=[\exp (\hbar
\omega _{m}/k_{B}T)-1]^{-1}$. $k_{B}$ is the Boltzmann constant and $T$ is
the temperature of the reservoir at the thermal equilibrium. We assume that $%
n_{\mathrm{th}}\equiv n_{q,\mathrm{th}}=n_{m,\mathrm{th}}$ for the resonant
condition $\omega _{0}=\omega _{m}$.

Before the numerical calculations of the correlation function of the
phonons, it is instructive to find the optimal conditions for strong phonon
antibunching. It has been shown in Ref.~\cite{RamosPRL13} that one can
observe the phonon antibunching (CPNB) at $\Delta =\pm J$ with strong
NAMR-qubit coupling ($J>\kappa ,\gamma $) at a low temperature.

Under the weak driving condition $\{\Omega,\varepsilon\}<\{\kappa,\gamma\}$, the optimal condition for UCPNB can be derived analytically following the approach given in Ref.~\cite{BambaPRA11}.
The optimal condition for UCPNB,
in the limit $T\rightarrow 0$, is a second-order equation in the strength ratio $\eta \equiv\Omega /\varepsilon $ as
\begin{equation}
A_{2}\eta ^{2}e^{-i2\phi }+A_{1}\eta e^{-i\phi }+A_{0}=0,  \label{Eq19}
\end{equation}%
with the coefficients $A_{i}$ ($i=0,1,2$) defined by
\begin{eqnarray}
A_{2} & \equiv &\sqrt{2}J_{\mathrm{opt}}^{2}, \label{Eqa6}\\
A_{1} & \equiv &-2\sqrt{2}J_{\mathrm{opt}}\left( 2\Delta _{\mathrm{opt}}-i\frac{%
\kappa +\gamma }{2}\right) , \label{Eqa7}\\
A_{0} & \equiv &\sqrt{2}J_{\mathrm{opt}}^{2}+\sqrt{2}\left( \Delta _{\mathrm{opt}}-i%
\frac{\kappa }{2}\right) \left( 2\Delta _{\mathrm{opt}}-i\frac{\kappa
+\gamma }{2}\right),\label{Eqa8}
\end{eqnarray}
where $\Delta _{\mathrm{opt}}$ and $J_{\mathrm{opt}}$ are the optimal
parameters for phonon blockade (i.e., $g_{b}^{\left( 2\right) }\left(
0\right) \rightarrow 0$) with the corresponding parameters of the external
driving fields ($\varepsilon $, $\Omega $ and $\phi $).
The details of the derivation for Eq.~(\ref{Eq19}) are given in Appendix~\ref{OpCo}.

If the microwave driving field is not applied to the qubit (i.e., $\Omega =0$%
), Eq.~(\ref{Eq19}) becomes
\begin{equation}
A_{0}=0,
\end{equation}%
and the optimal conditions are written as
\begin{eqnarray}
\Delta _{\mathrm{opt}} &=&0,  \label{Eq24} \\
J_{\mathrm{opt}} &=&\frac{1}{2}\sqrt{\kappa \left( \kappa +\gamma \right) }.
\label{Eq25}
\end{eqnarray}%
So if the microwave driving field is not applied to the qubit, the
parameters $\Delta _{\mathrm{opt}}$ and $J_{\mathrm{opt}}$ for phonon
blockade are determined by the decay rates ($\kappa $ and $\gamma $) of the
qubit and NAMR. For the case that a microwave driving field is applied to
the qubit (i.e., $\Omega \neq 0$), the two solutions of Eq.~(\ref{Eq19}) are
given by
\begin{eqnarray}
\eta _{\pm }e^{-i\phi _{\pm }} &=&\left( \frac{2\Delta _{\mathrm{opt}}}{J_{%
\mathrm{opt}}}-i\frac{\kappa +\gamma }{2J_{\mathrm{opt}}}\right)
\label{Eq26} \\
&\pm &\sqrt{\left( \frac{\Delta _{\mathrm{opt}}}{J_{\mathrm{opt}}}-i\frac{%
\gamma }{2J_{\mathrm{opt}}}\right) \left( \frac{2\Delta _{\mathrm{opt}}}{J_{%
\mathrm{opt}}}-i\frac{\kappa +\gamma }{2J_{\mathrm{opt}}}\right) -1}.  \notag
\end{eqnarray}%
This implies that we can choose the parameters $\Delta _{\mathrm{opt}}$ and $%
J_{\mathrm{opt}}$ for phonon blockade, and the corresponding parameters for
the two external driving fields ($\varepsilon $, $\Omega $ and $\phi $) are
determined by Eq.~(\ref{Eq26}).

\section{Phonon blockade without driving field applying on the qubit ($%
\Omega=0$)}

We now study the phonon statistics via the second-order correlation function $%
g_{b}^{\left( 2\right) }\left( \tau \right)$ and the mean phonon number $%
n_{b}=\left\langle b^{\dag }\left( t\right) b\left( t\right)\right\rangle $
when the driving field is not applied to the qubit (i.e., $\Omega =0$). They
are calculated numerically by solving the master equation [Eq.~(\ref{Eq6})]
within a truncated Fock space~\cite{LiewPRL10,MiranowiczPRA16}. The effect
of a microwave driving field applied to the qubit (i.e., $\Omega \neq 0$)
will be discussed in the next section.

\begin{figure}[tbp]
\includegraphics[bb=96 265 495 567, width=4.2 cm, clip]{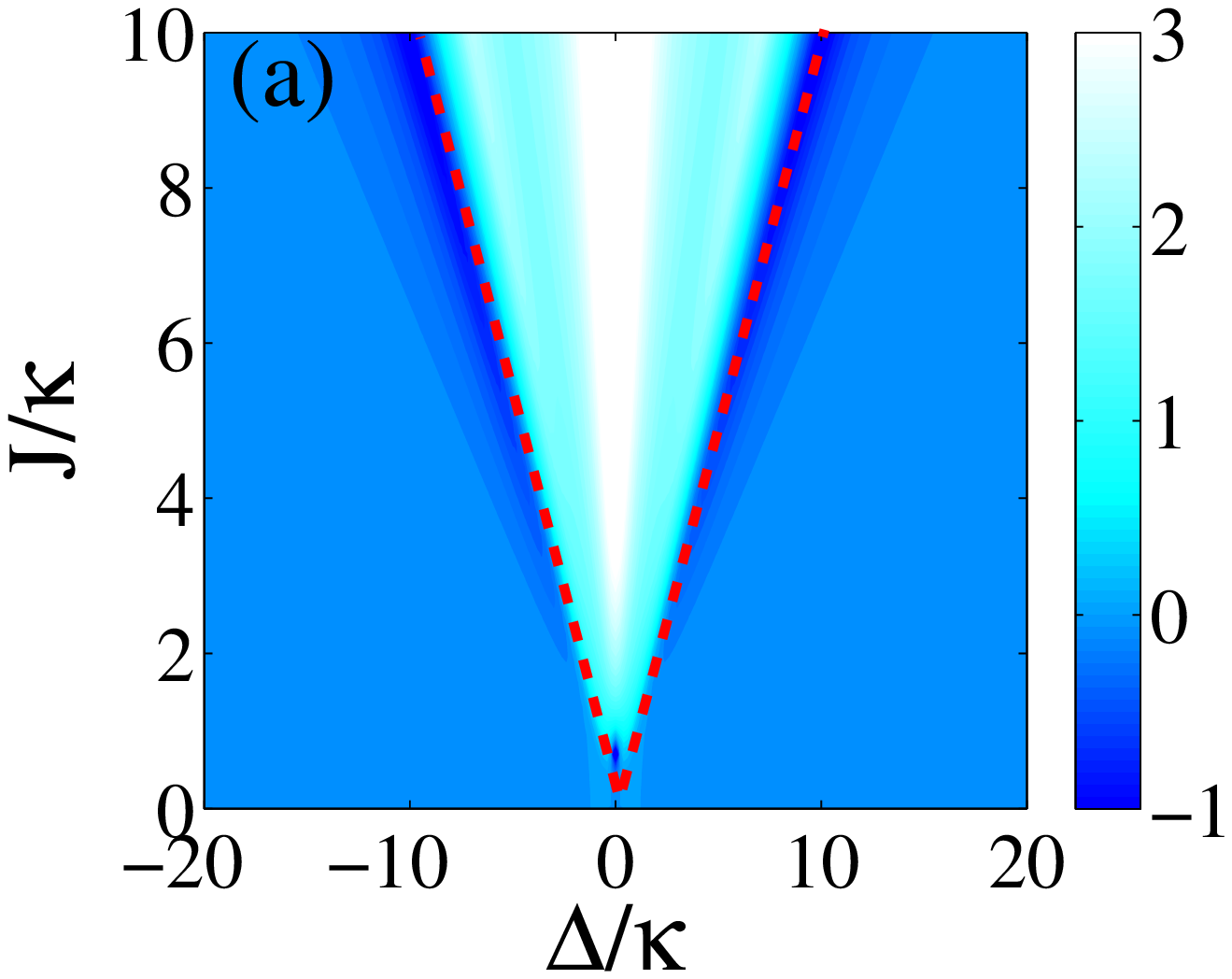} %
\includegraphics[bb=96 265 495 567, width=4.2 cm, clip]{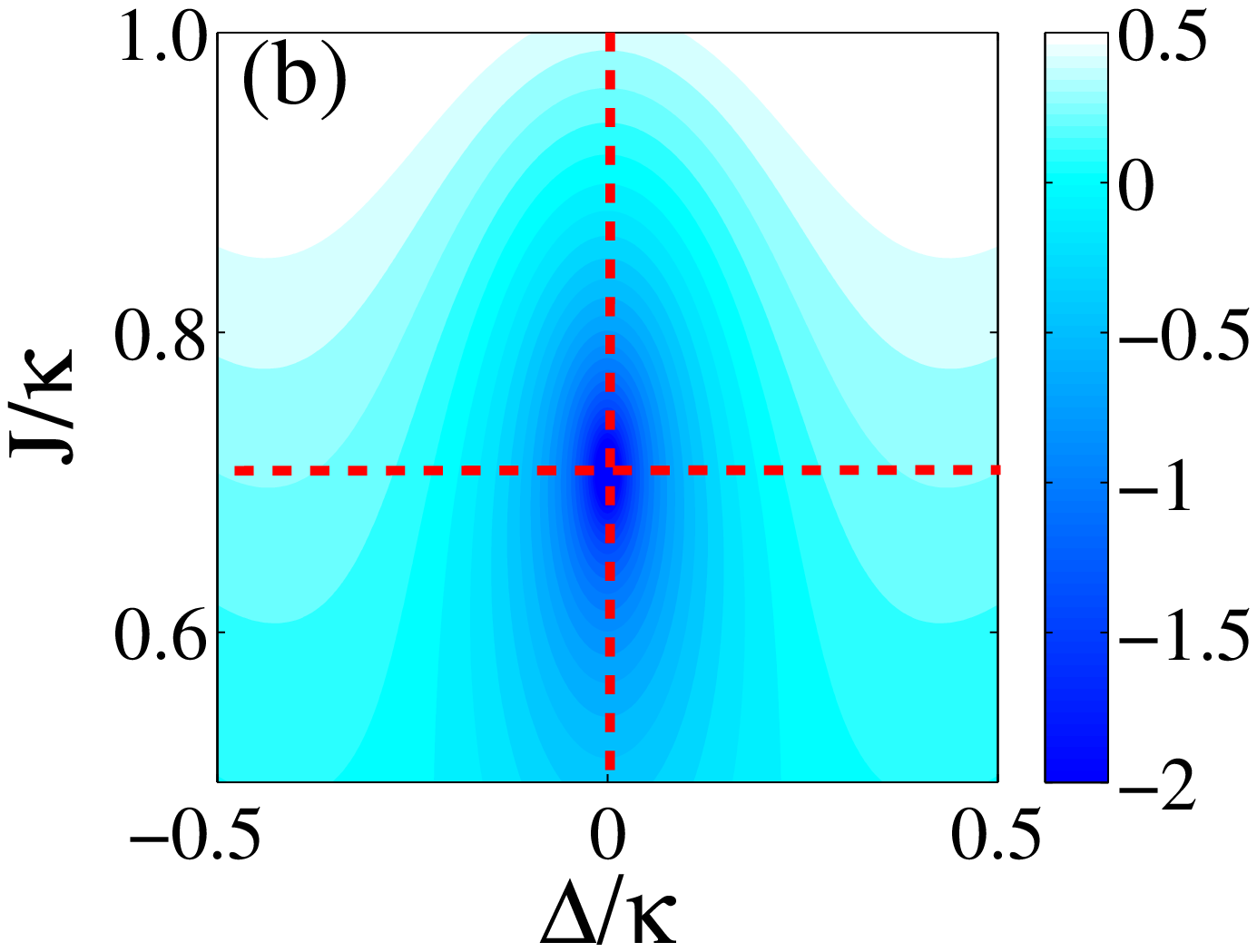} %
\includegraphics[bb=1 293 590 566, width=8.4 cm, clip]{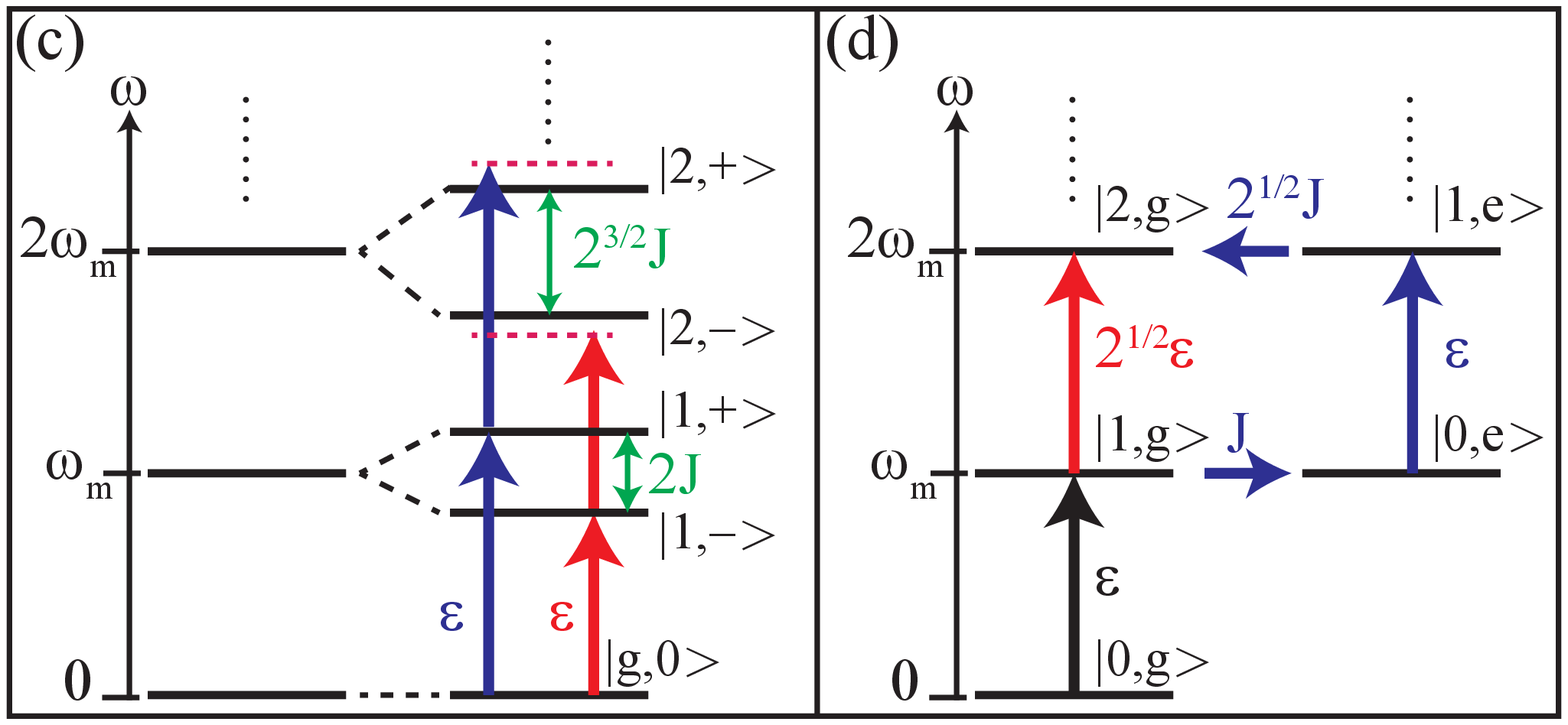}
\caption{(Color online) (a) Logarithmic plot (of base $10$) of the
equal-time second-order correlation function $\log_{10}g_{b}^{\left(
2\right) }\left( 0\right)$ as a function of the detuning $\Delta/\protect%
\kappa$ and the linear coupling strength $J/\protect\kappa$ for the
parameters: $\protect\gamma=\protect\kappa$, $\protect\varepsilon=0.01%
\protect\kappa$, $\Omega=0$ and $n_{\mathrm{th}}=0$. (b) shows a magnification of (a) at small $\Delta$ and $J$. (c) Energy-level
diagram of one mechanical mode strongly coupled to a qubit. (d) Transition
paths leading to the quantum interference responsible for the strong phonon
antibunching. }
\label{fig2}
\end{figure}

In Fig.~\ref{fig2}, the equal-time second-order correlation function $%
g_{b}^{\left( 2\right) }\left( 0\right) $ is plotted as a function of the
rescaled detuning $\Delta /\kappa $ and the rescaled coupling strength $%
J/\kappa $. Phonon blockade appears in two different parameter regions: (i) $%
\Delta =\pm J$ for $J>\kappa $, i.e., the blue (dark) region along with the
red dash lines in Fig.~\ref{fig2}(a); (ii) the blue (dark) region around the
point ($\Delta =0$, $J=\kappa /\sqrt{2}\approx 0.71\kappa$), as shown in
Fig.~\ref{fig2}(b). 

Phonon blockade appearing in two different parameter regions can be
interpreted in two different ways. Phonon blockade appearing in the region
(i) ($\Delta =\pm J$ for $J>\kappa$) results from the anharmonicity of the
eigenstates of the strong coupling NAMR-qubit system. The energy-level
diagram of one mechanical mode strongly coupled to a qubit is shown in Fig.~%
\ref{fig2}(c), where $|n, \pm\rangle$ are the eigenstates (dressed states)
of a mechanical mode (with the phonon number states $|n\rangle$) coupled to
a qubit (with two states $|g\rangle$ and $|e\rangle$). The phonon, absorbed
resonantly by the transition from $|0, g\rangle$ to $|1, \pm\rangle$, blocks
the transition from $|1, g\rangle$ to $|2, \pm\rangle$ for large detuning.
This kind of phonon blockade can be called as conventional phonon blockade
(CPNB) in analogy with the phenomena of photon blockade in cavity QED~\cite%
{BirnbaumNat05,FaraonNPy08,CLangPRL11,HoffmanPRL11}.

Phonon blockade appearing in the region (ii) with weak coupling strength ($%
J<\kappa $) can be explained using the destructive interference between two
different paths of two-phonon excitation as illustrated in Fig.~\ref{fig2}%
(d) for transition paths $\left\vert 1,g\right\rangle \rightarrow \left\vert
2,g\right\rangle $ and $\left\vert 1,g\right\rangle \rightarrow \left\vert
0,e\right\rangle \rightarrow \left\vert 1,e\right\rangle \rightarrow
\left\vert 2,g\right\rangle$. We call the interference-based
phonon blockade appearing in the region (ii) as unconventional phonon
blockade (UCPNB). It is worth mentioning that the strong photon antibunching based on quantum interference has been found in the weakly nonlinear photonic molecules~\cite{LiewPRL10,BambaPRA11,LemondePRA14,MajumdarPRL12,KomarPRA13,GeracePRA14,KyriienkoPRA14a,XuPRA14a,XuPRA14b,KyriienkoPRA14,ShenPRA15,ZhouPRA15}.
However, their is difference between our study here and that of nonlinear photonic molecules. The main difference is that there are two separate energy scales in the weakly nonlinear photonic molecules~\cite{LiewPRL10,BambaPRA11,LemondePRA14,MajumdarPRL12,KomarPRA13,GeracePRA14,KyriienkoPRA14a,XuPRA14a,XuPRA14b,KyriienkoPRA14,ShenPRA15,ZhouPRA15},
one is large linear coupling strength between coupled cavity modes, and second is small nonlinearity (up
to hundred times smaller than the photon damping rate). In the NAMR-qubit system, there is only a single energy parameter $J$, i.e., the NAMR-qubit coupling strength, which is reminiscent to the tunneling rate in photonic systems and is the order of the damping rates of the qubit or NAMR [Eq.~(\ref{Eq25})], while a large
nonlinearity is hidden in the qubit structure.

The conditions for the appearance of phonon blockade in region (ii) can be understand by the following way. For $\Delta=0$, the amplitude for transition $\left\vert 1,g\right\rangle \rightarrow \left\vert
2,g\right\rangle $ is proportional to $\sqrt{2}\varepsilon$ and the amplitude for  transition $\left\vert 1,g\right\rangle \rightarrow \left\vert0,e\right\rangle \rightarrow \left\vert 1,e\right\rangle \rightarrow
\left\vert 2,g\right\rangle$ is proportional to $4\sqrt{2}\varepsilon J^{2}/[\kappa(\kappa+\gamma)]$, where $\kappa/2$ and $(\kappa+\gamma)/2$ are the decay rates of the states $\left\vert
0,e\right\rangle$ and $\left\vert 1,e\right\rangle$, respectively. The perfect destructive quantum interference for strong phonon antibunching is only achieved in the condition that the transition matrix elements of these two paths of phonon excitation have the same amplitude but inverse phase, thus we obtain the optimal condition $J_{\rm opt} \propto \sqrt{\kappa(\kappa+\gamma)}/2$. This is consistent well with the analytical result given in Eqs.~(\ref{Eq25}).

\begin{figure}[tbp]
\includegraphics[bb=6 111 589 761, width=8.5 cm, clip]{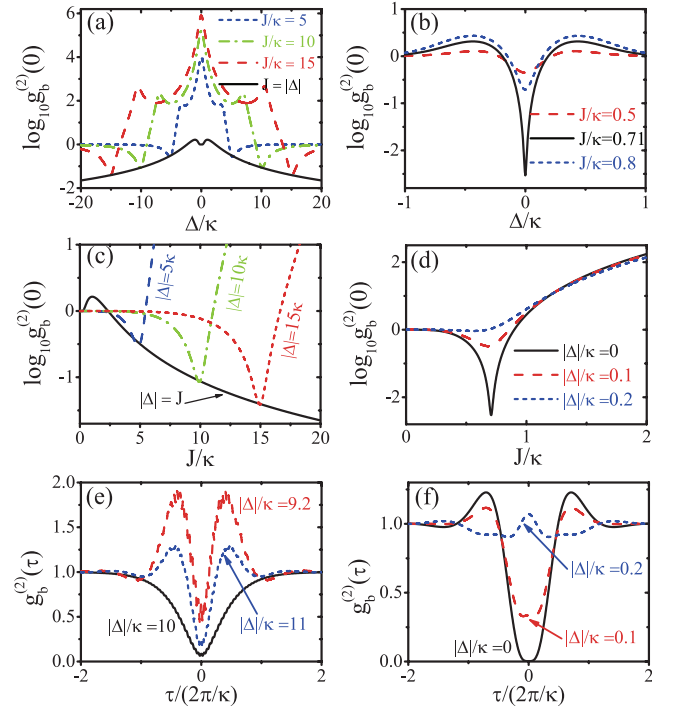}
\caption{(Color online) The equal-time second-order correlation function $%
\log_{10}g_{b}^{\left( 2\right) }\left( 0\right)$ is plotted as a function of
the detuning $\Delta/\protect\kappa$ in (a) for $J/\kappa$ taking the values $5$, $10$, $15$, $|\Delta|/\kappa$ and in
(b) for $J/\kappa$ taking the values $0.5$, $0.71$, $0.8$, and as a function of the coupling strength $%
J/\protect\kappa$ in (c) for $|\Delta|/\kappa$ taking the values $5$, $10$, $15$, $J/\kappa$ and in (d) for $|\Delta|/\kappa$ taking the values $0$, $0.1$, $0.2$. The second-order correlation function $g_{b}^{\left( 2\right) }\left( \protect%
\tau\right)$ is plotted as a function of the normalized time delay $\protect\tau%
/(2\protect\pi/\protect\kappa)$ in (e) for $J=10\protect\kappa$ and $|\Delta|/\kappa$ taking the values $10$, $9.2$, $11$ and in (f) for $J/\protect\kappa=0.71$ and $|\Delta|/\kappa$ taking the values $0$, $0.1$, $0.2$. The other parameters are taken as $
\protect\gamma=\protect\kappa$, $\protect\varepsilon=0.01\protect\kappa$, $%
\Omega=0$ and $n_{\mathrm{th}}=0$.}
\label{fig3}
\end{figure}

In order to see these more clearly, a few snapshots taken from Figs.~\ref%
{fig2}(a) and \ref{fig2}(b) are shown in Fig.~\ref{fig3}. The equal-time
second-order correlation functions $g_{b}^{\left( 2\right) }\left( 0\right) $
are plotted as functions of the rescaled detuning $\Delta /\kappa $ in Figs.~%
\ref{fig3}(a) and \ref{fig3}(b) and plotted as the function of the rescaled
coupling strength $J/\kappa $ in Figs.~\ref{fig3}(c) and \ref{fig3}(d). As
shown in Fig.~\ref{fig3}(a) for CPNB, with a strong NAMR-qubit coupling
strength $J=10 \kappa$, we can obtain $g_{b}^{\left( 2\right) }\left(
0\right) \approx 0.1 $ at $\Delta=\pm J$, and CPNB should be more easily
observed for larger NAMR-qubit coupling strength $J$ [see, Fig.~\ref{fig3}(c)].
As illustrated in Figs.~\ref{fig3}(b) and \ref{fig3}(d), the minimal value of the second-order correlation function $g_{b}^{\left( 2\right) }\left( 0\right)\approx 0.003$ can be reached with a weak NAMR-qubit coupling strength ($J=\kappa/\sqrt{2}$) at $\Delta=0 $ [predicted by Eqs.~(\ref{Eq24}) and (\ref{Eq25}) for UCPNB]. The second-order
correlation function $g_{b}^{\left( 2\right) }\left( \tau \right) $ is
plotted as a function of the normalized time delay $\tau /(2\pi /\kappa )$
in Fig.~\ref{fig3}(e) for $J=10\protect\kappa$ (CPNB) and $|\Delta|/\kappa$ taking the values $10$, $9.2$, $11$ and in Fig.~\ref%
{fig3}(f) for $J/\protect\kappa=0.71$ (UCPNB) and $\Delta/\kappa$ taking the values $0$, $0.1$, $0.2$.
The time duration for both CPNB and UCPNB is of the order of the life time of the phonons in the NAMR.

\begin{figure}[tbp]
\includegraphics[bb=5 221 580 633, width=8.5 cm, clip]{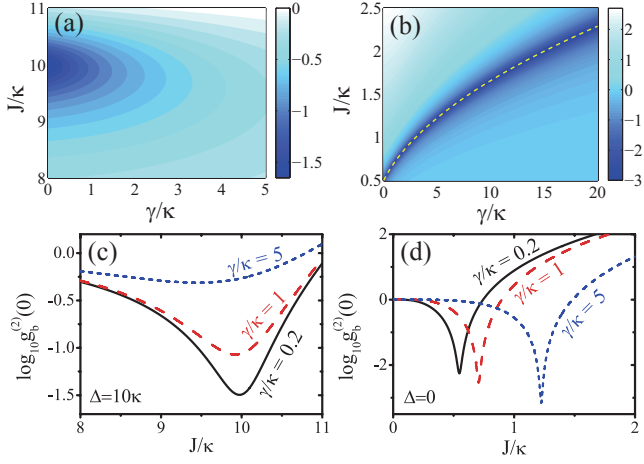}
\caption{(Color online)
Logarithmic plot of the
equal-time second-order correlation function $\log_{10}g_{b}^{\left(
2\right) }\left( 0\right)$ as functions of $\gamma/\protect%
\kappa$ and $J/\protect\kappa$ in (a) for $\Delta=10\kappa$ and in (b) for $\Delta=0$.
Also $\log_{10}g_{b}^{\left( 2\right) }\left( 0\right)$ plotted as a function of $J/\kappa$ in (c) for $\Delta=10\kappa$ and in (d) for $\Delta=0$ with $\gamma/\kappa=0.2$, $1$, and $5$.
The other parameters are taken as $\varepsilon=0.01%
\protect\kappa$, $\Omega=0$ and $n_{\mathrm{th}}=0$. The yellow dash curve in (b) denotes the optimal condition of $J$ given in Eq.~(\ref{Eq25}).}
\label{fig4}
\end{figure}

Figures~\ref{fig4}(a) and \ref{fig4}(b) display the equal-time second-order correlation function $
\log_{10}g_{b}^{\left( 2\right) }\left( 0\right)$ as functions of $\gamma/\protect%
\kappa$ and $J/\protect\kappa$ for $\Delta=10\kappa$ in Fig.~\ref{fig4}(a) and  $\Delta=0$ in Fig.~\ref{fig4}(b). $\log_{10}g_{b}^{\left( 2\right) }\left( 0\right)$ as a function of $J/\kappa$ with different $\gamma/\kappa$ is shown in Fig.~\ref{fig4}(c) for $\Delta=10\kappa$ and in Fig.~\ref{fig4}(d) for $\Delta=0$. With increasing $\gamma/\kappa$, CPNB at $\Delta=10\kappa$ has been gradually weakened [see, Figs.~\ref{fig4}(a) and \ref{fig4}(c)] and a larger value of $J/\kappa$ is need for the same value of $g^{(2)}_b(0)$. Similarly, the optimal value of $J/\kappa$ for UCPNB increases with $\gamma/\kappa$ [see, Figs.~\ref{fig4}(b) and \ref{fig4}(d)]. The optimal value of $J/\kappa$ for UCPNB [blue dark regime in Fig.~\ref{fig4}(b)] is agree well with the analytical result given in Eq.~(\ref{Eq25}) [yellow dash curve in Fig.~\ref{fig4}(b)]. For simplicity of discussions, we choose $\gamma=\kappa$ in the following. However, we should mention that different ratios $\gamma/\kappa$ corresponded to different optimal conditions for UCPNB.

Different from the study on photon blockade in cavity QED or nonlinear
photonic molecules, where the thermal photon effects can be neglected
safely, because the mean thermal photon number for an optical cavity is
negligible small, e.g., the thermal photon number for an optical cavity with
frequency $3.21\times 10^{14}$ Hz in Ref.~\cite{{BirnbaumNat05}} is smaller
than $10^{-22}$ at the room temperature. While the effect of the thermal
phonons should be considered in the observation of phonon blockade even with
microwave-frequency nanomechanical resonators. Because the mean thermal
phonon number becomes about $10^{-5}$ for a mechanical resonator with
resonance frequency of $6$~GHz at a temperature of $25$~mK~\cite%
{OConnellNat10} and we will find that even this small mean thermal phonon
number may have significant effect on the observation of phonon blockade.

\begin{figure}[tbp]
\includegraphics[bb=14 239 581 643, width=8.5 cm, clip]{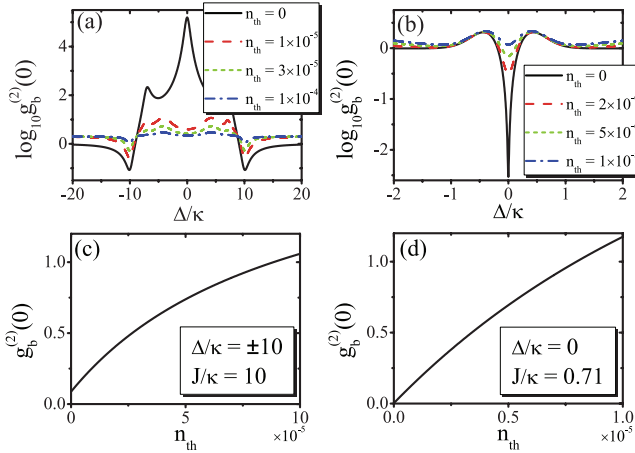}
\caption{(Color online) The equal-time second-order correlation functions $%
\log_{10}g_{b}^{\left( 2\right) }\left( 0\right)$ is plotted as a function of
the detuning $\Delta/\protect\kappa$ in (a) for $J/\protect\kappa=10$ and in
(b) for $J/\protect\kappa=0.71$ with different mean thermal photon (phonon)
number $n_{\mathrm{th}}$. The equal-time second-order correlation functions $%
g_{b}^{\left( 2\right) }\left( 0\right)$ is plotted as a function of the mean
thermal photon (phonon) number $n_{\mathrm{th}}$ in (c) for $J=|\Delta|=10/%
\protect\kappa$ and in (d) for $J/\protect\kappa=0.71$ and $\Delta=0$. The
other parameters are taken as $\protect\gamma=\protect\kappa$, $\protect\varepsilon%
=0.01\protect\kappa$, and $\Omega=0$.}
\label{fig5}
\end{figure}

We present in Figs.~\ref{fig5}(a) and \ref{fig5}(b) the equal-time
second-order correlation functions $g_{b}^{\left( 2\right) }\left( 0\right)$
as functions of the rescaled detuning $\Delta/\kappa$ with different mean thermal
photon (phonon) number $n_{\mathrm{th}}$. The equal-time second-order
correlation functions $g_{b}^{\left( 2\right) }\left( 0\right)$ are plotted
as a function of the mean thermal photon (phonon) number $n_{\mathrm{th}}$ in
Figs.~\ref{fig5}(c) and \ref{fig5}(d). It is clear that the thermal phonons
(photons) have a significant effect on both the CPNB and UCPNB under weak
driving condition ($\varepsilon=0.01\kappa$). The CPNB disappears when the
number of the mean thermal phonons reaches $n_{\mathrm{th}}=8.85\times 10^{-5}$;
the UCPNB disappears for the mean number of the thermal phonons $n_{\mathrm{th}%
}=0.8\times 10^{-5}$. So that means the UCPNB is more fragile against the
thermal noise than the CPNB.

\begin{figure}[tbp]
\includegraphics[bb=6 206 590 631, width=8.5 cm, clip]{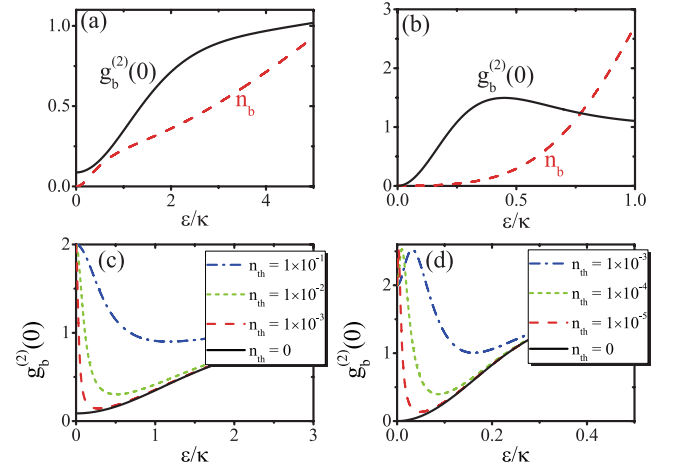}
\caption{(Color online) The equal-time second-order correlation functions $%
g_{b}^{\left( 2\right) }\left( 0\right)$ (solid line) and mean phonon number
$n_{b}$ (dashed line) are plotted as functions of the mechanical driving
strength $\protect\varepsilon/\protect\kappa$ in (a) for $J=|\Delta|=10/%
\protect\kappa$ and in (b) for $J/\protect\kappa=0.71$ and $\Delta=0$. $%
g_{b}^{\left( 2\right) }\left( 0\right)$ is plotted as a function of the
mechanical driving strength $\protect\varepsilon/\protect\kappa$ in (c) for $%
J=|\Delta|=10/\protect\kappa$ and in (d) for $J/\protect\kappa=0.71$ and $%
\Delta=0$ with different thermal photon (phonon) number $n_{\mathrm{th}}$.
The other parameters are taken as $\protect\gamma=\protect\kappa$, $\protect%
\varepsilon=0.01\protect\kappa$, and $\Omega=0$.}
\label{fig6}
\end{figure}

One of the main reasons for the CPNB and UCPNB disappearing in such a small mean
thermal photon (phonon) number ($n_{\mathrm{th}}<1\times 10^{-4}$) is that
the (total) mean phonon number $n_{b}$ in the NAMR is very small ($n_{b}
\sim 1\times 10^{-4}$) for weak external driving ($\varepsilon=0.01\kappa$).
So in order to improve the robustness against the thermal phonons, we need a
larger number of mean phonons in antibunching and one simple way is to
enhance the driving strength of the external field $\varepsilon$. The
equal-time second-order correlation functions $g_{b}^{\left( 2\right)
}\left( 0\right)$ and mean phonon number $n_{b}$ are plotted as functions of
the mechanical driving strength $\varepsilon/\kappa$ in Figs.~\ref{fig6}(a)
and \ref{fig6}(b), respectively. The mean phonon number $n_{b}$ increases as
enhancing the driving strength, while the phonons tend to behave classically
($g_{b}^{\left( 2\right) }\left( 0\right)>1$) when the driving strength
becomes strong enough. What's more, we can see that the $n_{b}$ in UCPNB is
much smaller than the one in CPNB for the same value of $g_{b}^{\left(
2\right) }\left( 0\right)$. As shown in Figs.~\ref{fig6}(c) and \ref{fig6}%
(d), the UCPNB can only be observed in much smaller mean number of the thermal
phonons than the one for CPNB.

According to the results of Fig.~\ref{fig2} to Fig.~\ref{fig6}, we now
summarize the properties of CPNB and UCPNB: UCPNB appears in the weak
NAMR-qubit coupling condition and the phonon blockade is nearly perfect ($%
g_{b}^{\left( 2\right) }\left( 0\right) < 0.01$), but UCPNB can only be
observed with a small mean phonon number and it is very fragile in a
reservoir with thermal noise; in contrast to UCPNB, CPNB can be observed
with much larger mean phonon number and is more robust against the thermal
noise, but CPNB needs very strong NAMR-qubit coupling for strong phonon
antibunching. Another reason for us to enlarge the mean phonon number for
strong phonon antibunching is that we need to increase the number of single
phonons generated by a single-phonon source in a given time.

\section{Phonon blockade with a driving field applying to the qubit $%
\Omega\neq0$}

\begin{figure}[tbp]
\includegraphics[bb=45 338 567 561, width=8.5 cm, clip]{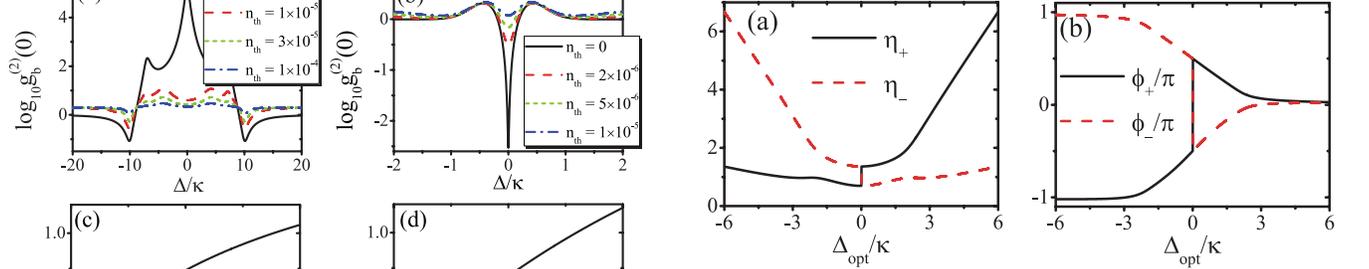}
\caption{(Color online) (a) $\protect\eta_{\pm}/\protect\epsilon$ and (b) $%
\protect\phi_{\pm}/\protect\pi$ from Eq.~(\protect\ref{Eq26}) are plotted as
functions of the frequency of $\Delta_{\mathrm{opt}}/\protect\kappa$. The
other parameters are taken as $\protect\gamma=\protect\kappa$ and $J=J_{\mathrm{opt}%
}=3\protect\kappa$.}
\label{fig7}
\end{figure}

\begin{figure}[tbp]
\includegraphics[bb=81 265 489 568, width=4.2 cm, clip]{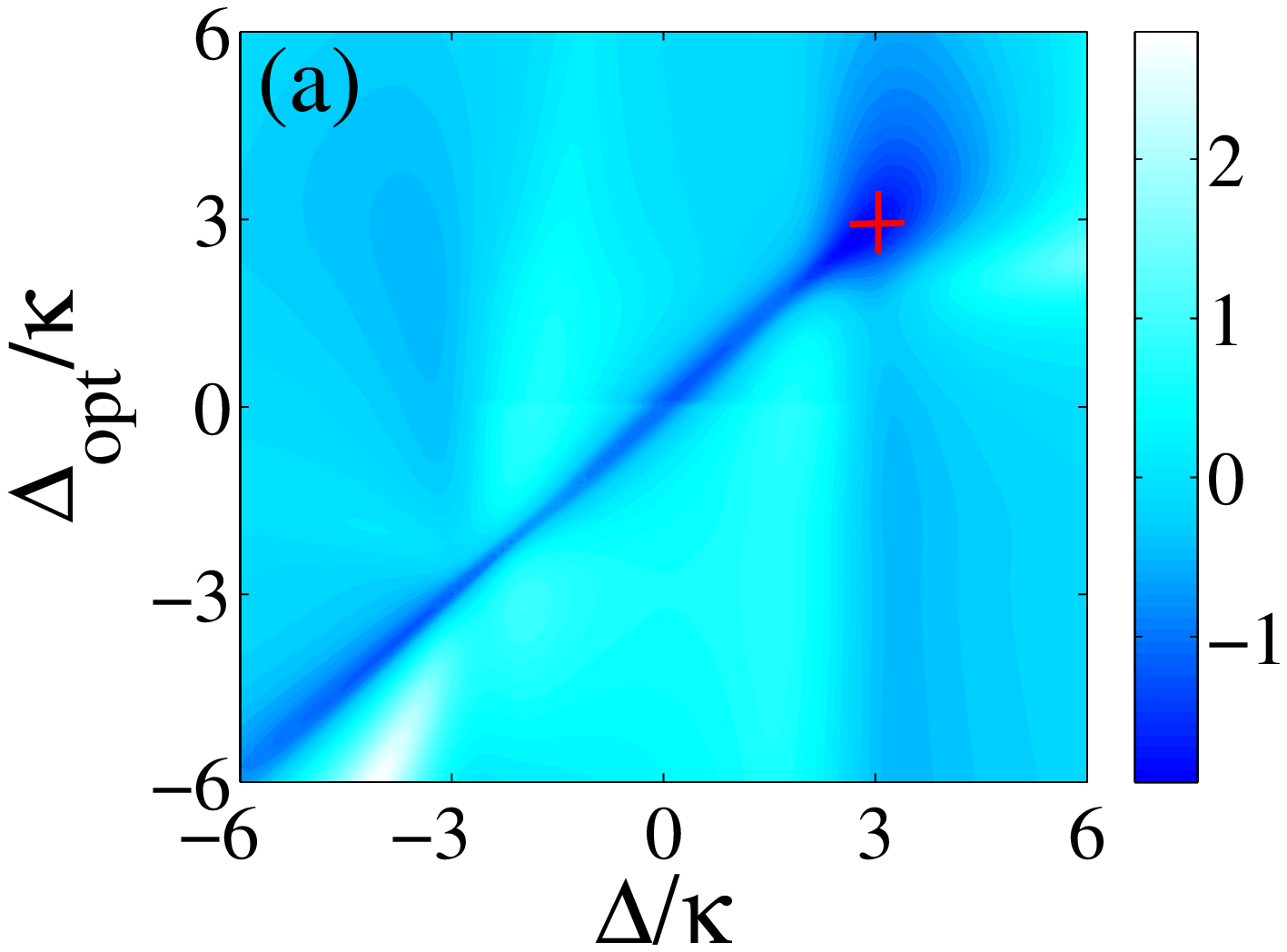} %
\includegraphics[bb=81 265 489 568, width=4.2 cm, clip]{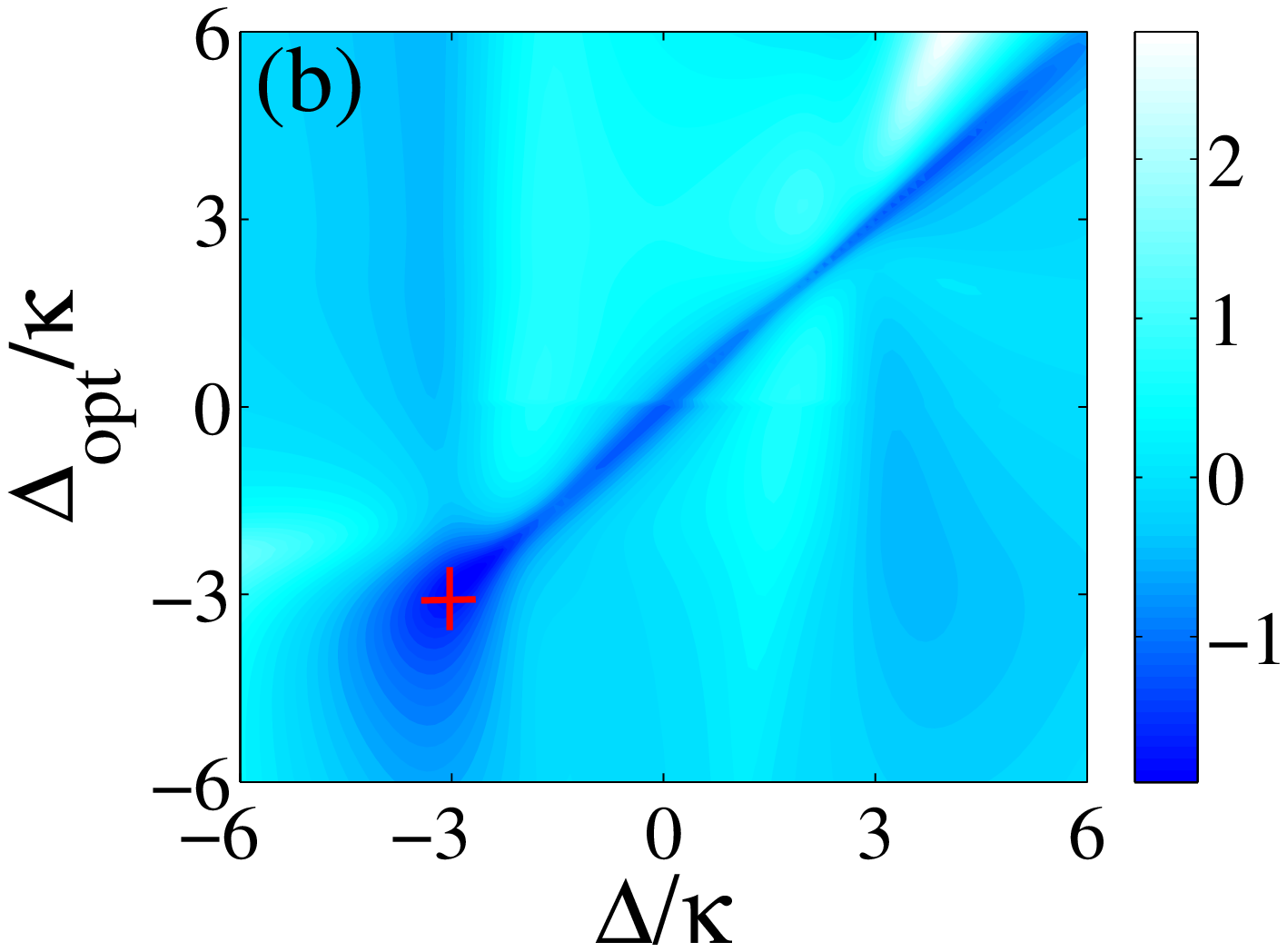}
\caption{(Color online) Logarithmic plot of the equal-time
second-order correlation functions $\log_{10}g_{b}^{\left( 2\right) }\left(
0\right)$ as functions of the detuning $\Delta/\protect\kappa$ and $\Delta_{%
\mathrm{opt}}/\protect\kappa$ in (a) for $\protect\eta=\protect\eta_{+}$ and
$\protect\phi=\protect\phi_{+}$ and in (b) for $\protect\eta=\protect\eta%
_{-} $ and $\protect\phi=\protect\phi_{-}$. The red cross marked at (a) $%
\Delta=\Delta_{\mathrm{opt}}\approx 3\protect\kappa$ and (b) $\Delta=\Delta_{%
\mathrm{opt}}\approx -3\protect\kappa$ denotes the minimal point of $%
g_{b}^{\left( 2\right) }\left(0\right)$. The other parameters are taken as $\protect%
\gamma=\protect\kappa$, $J=J_{\mathrm{opt}}=3\protect\kappa$, $\protect%
\varepsilon=\protect\kappa/5$, and $n_{\mathrm{th}}=0$.}
\label{fig8}
\end{figure}

We now discuss how to enlarge the mean phonon number for strong phonon
antibunching with a moderate NAMR-qubit coupling strength ($J=J_{\mathrm{opt}%
}=3 \kappa$) by applying an external driving field to the qubit (i.e., $%
\Omega \neq 0$). As shown in Eq.~(\ref{Eq26}), the optimal parameters ($%
\Delta_{\mathrm{opt}}$ and $J_{\mathrm{opt}}$) for UCPNB is related to the
strength ratio and phase difference ($\eta_{\pm}$ and $\phi_{\pm}$) of the
driving fields, so we can combine CPNB and UCPNB together (appear in the
same parameter region, i.e., $\Delta_{\mathrm{opt}}=\pm J$) by optimizing
the values of $\eta_{\pm}$ and $\phi_{\pm}$. In Fig.~\ref{fig7}, parameters $%
\eta_{\pm}$ and $\phi_{\pm}/\pi$ in Eq.~(\ref{Eq26}) are plotted as
functions of the detuning of $\Delta_{\mathrm{opt}}/\kappa$. We find that
there is an abrupt change of $\eta_{\pm}$ and $\phi_{\pm}$ at $\Delta_{%
\mathrm{opt}}=0$, and we also find that $\eta_{+}>\eta_{-}$ in the region of
$\Delta_{\mathrm{opt}}>0$ and $\eta_{+}<\eta_{-}$ for $\Delta_{\mathrm{opt}%
}<0$.

To compare the two solutions in Eq.~(\ref{Eq26}) (optimal conditions) for
phonon blockade, the equal-time second-order correlation function $%
g_{b}^{\left( 2\right) }\left(0\right)$ is shown as a function of the
rescaled detunings $\Delta/\kappa$ and $\Delta_{\mathrm{opt}}/\kappa$ in
Fig.~\ref{fig8}(a) for $(\eta,\phi)=(\eta_{+},\phi_{+})$ and in \ref{fig8}%
(b) for $(\eta,\phi)=(\eta_{-},\phi_{-})$. Phonons exhibit strong
antibunching (blue or dark) in the region around the line $\Delta=\Delta_{%
\mathrm{opt}}$, and the minimal value of $g_{b}^{\left( 2\right)
}\left(0\right)$ appears at $\Delta=\Delta_{\mathrm{opt}}\approx J$ (red
cross mark) in Fig.~\ref{fig8}(a) for $(\eta,\phi)=(\phi_{+},\eta_{+})$ and
at $\Delta=\Delta_{\mathrm{opt}}\approx -J$ (red cross mark) in Fig.~\ref%
{fig8}(b) for $(\eta,\phi)=(\phi_{-},\eta_{-})$.

\begin{widetext}
\begin{figure*}[tbp]
\includegraphics[bb=10 339 579 625, width=16.5 cm, clip]{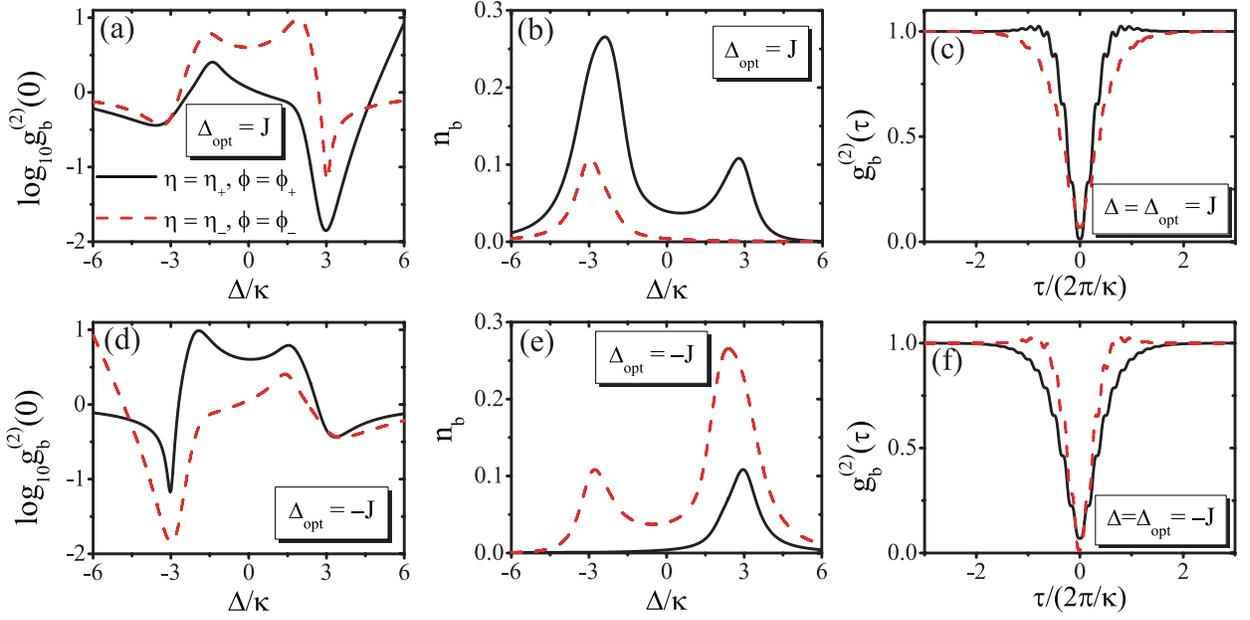}
\caption{(Color online) The equal-time second-order correlation functions $\log_{10}g_{b}^{\left( 2\right) }\left( 0\right)$ [(a) and (d)] and mean phonon number $n_{b}$ [(b) and (e)] are plotted as functions of the detuning $\Delta/\kappa$; $g_{b}^{\left( 2\right) }\left( \tau\right)$ is plotted as a function of the normalized time delay $\tau/(2\pi/\kappa)$ in (c) and (f) [(a), (b) and (c) for $\Delta_{\rm opt}=J$; (d), (e) and (f) for $\Delta_{\rm opt}=-J$]. The solid line is plotted for $(\eta,\phi)=(\eta_{+},\phi_{+})$ and the dashed line is plotted for $(\eta,\phi)=(\eta_{-},\phi_{-})$. The other parameters are taken as $\gamma=\kappa$, $\varepsilon=\kappa/5$, $J=J_{\rm opt}=3\kappa$, and $n_{\rm th}=0$.}
\label{fig9}
\end{figure*}
\end{widetext}

Two snapshots (along the lines of $\Delta_{\mathrm{opt}}=\pm J$) taken from
Figs.~\ref{fig8}(a) and \ref{fig8}(b) are shown in Figs.~\ref{fig9}(a) and %
\ref{fig9}(d), and the corresponding mean phonon number $n_{b}$ is shown in
Figs.~\ref{fig9}(b) and \ref{fig9}(e). Fig.~\ref{fig9}(a) shows that $%
g_{b}^{\left( 2\right) }\left(0\right)$ for $(\eta,\phi)=(\eta_{+},\phi_{+})$
is smaller than the one for $(\eta,\phi)=(\eta_{-},\phi_{-})$ at $%
\Delta=J=3\kappa$ when $\Delta_{\mathrm{opt}}=J$; in contrast, when $\Delta_{%
\mathrm{opt}}=-J$ as shown in Fig.~\ref{fig9}(d), $g_{b}^{\left( 2\right)
}\left(0\right)$ for $(\eta,\phi)=(\eta_{+},\phi_{+})$ is larger than the
one for $(\eta,\phi)=(\eta_{-},\phi_{-})$ at $\Delta=-J=-3\kappa$.
Correspondingly, the mean phonon number $n_{b}$ for $(\eta,\phi)=(\eta_{+},%
\phi_{+})$ is much larger than the one for $(\eta,\phi)=(\eta_{-},\phi_{-})$
at $\Delta=J=3\kappa$ when $\Delta_{\mathrm{opt}}=J$ as shown in Fig.~\ref%
{fig9}(b); in contrast, $n_{b}$ for $(\eta,\phi)=(\eta_{+},\phi_{+})$ is
much smaller than the one for $(\eta,\phi)=(\eta_{-},\phi_{-})$ at $%
\Delta=-J=-3\kappa$ when $\Delta_{\mathrm{opt}}=-J$ as shown in Fig.~\ref%
{fig9}(e). More important, when $(J,\varepsilon)=(3\kappa,\kappa/5)$, the
mean phonon number $n_{b}$ can be larger than $0.1$ with the optimal
parameters $(\eta,\phi,\Delta,\Delta_{\mathrm{opt}})=(\eta_{+},\phi_{+},%
\Delta_{\mathrm{opt}},J)$ or $(\eta,\phi,\Delta,\Delta_{\mathrm{opt}%
})=(\eta_{-},\phi_{-},\Delta_{\mathrm{opt}},-J)$ for phonon blockade [$%
g_{b}^{\left( 2\right) }\left(0\right)\approx 0.0141$]. So in order to
obtain a phonon blockade with small $g_{b}^{\left( 2\right) }\left(0\right)$
and large mean phonon number, we should choose the parameters $%
(\eta,\phi,\Delta,\Delta_{\mathrm{opt}})=(\eta_{+},\phi_{+},\Delta_{\mathrm{%
opt}},J)$ or $(\eta,\phi,\Delta,\Delta_{\mathrm{opt}})=(\eta_{-},\phi_{-},%
\Delta_{\mathrm{opt}},-J)$. The second-order correlation $g_{b}^{\left(
2\right) }\left( \tau\right)$ is plotted as a function of the normalized
time delay $\tau/(2\pi/\kappa)$ in Figs.~\ref{fig9}(c) and \ref{fig9}(f).
The time durations of the correlated phonons are about the life time of the
phonons in the NAMR.

\begin{figure}[tbp]
\includegraphics[bb=10 346 586 564, width=8.5 cm, clip]{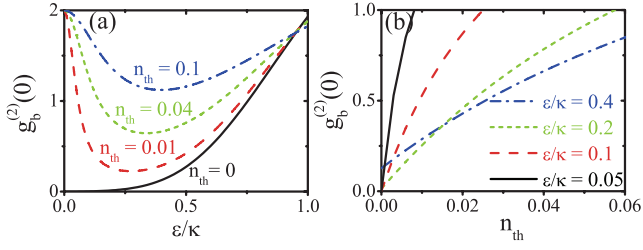}
\caption{(Color online) (a) The equal-time second-order correlation
functions $g_{b}^{\left( 2\right) }\left( 0\right)$ is plotted as a function of
the mechanical driving strength $\protect\varepsilon/\protect\kappa$ with
different mean thermal photon (phonon) number $n_{\mathrm{th}}$; (b) $%
g_{b}^{\left( 2\right) }\left( 0\right)$ is plotted as a function of the
mean thermal photon (phonon) number $n_{\mathrm{th}}$ with different mechanical
driving strength $\protect\varepsilon/\protect\kappa$ for $(\protect\eta,%
\protect\phi,\Delta,\Delta_{\mathrm{opt}})=(\protect\eta_{+},\protect\phi%
_{+},\Delta_{\mathrm{opt}},J)$ or $(\protect\eta,\protect\phi,\Delta,\Delta_{%
\mathrm{opt}})=(\protect\eta_{-},\protect\phi_{-},\Delta_{\mathrm{opt}},-J)$%
. The other parameters are taken as $\protect\gamma=\protect\kappa$ and $J=J_{\mathrm{%
opt}}=3\protect\kappa$.}
\label{fig10}
\end{figure}

Next, let us discuss the robustness of the phonon blockade effect against
the thermal noise with parameters $(\eta ,\phi ,\Delta ,\Delta _{\mathrm{opt}%
})=(\eta _{+},\phi _{+},\Delta _{\mathrm{opt}},J)$ or $(\eta ,\phi ,\Delta
,\Delta _{\mathrm{opt}})=(\eta _{-},\phi _{-},\Delta _{\mathrm{opt}},-J)$.
Figure~\ref{fig10}(a) shows the equal-time second-order correlation functions
$g_{b}^{\left( 2\right) }\left( 0\right) $ as a function of the mechanical
driving strength $\varepsilon /\kappa $ for different mean thermal photon
(phonon) number $n_{\mathrm{th}}$, and Fig.~\ref{fig10}(b) shows $%
g_{b}^{\left( 2\right) }\left( 0\right) $ as a function of the mean thermal
photon (phonon) number $n_{\mathrm{th}}$ for different mechanical driving
strength $\varepsilon /\kappa $. From Fig.~\ref{fig10}, we can see that the
phonon blockade can be observed with the thermal phonon number of $0.06$
when $\varepsilon =0.4\kappa $.

\section{measurements of phonon blockade}

We now turn to study the measurements of phonon blockade in the NAMR-qubit system via an optical cavity field. We assume that the
optical cavity field is coupled to the NAMR through the radiation pressure type interaction. We will show that the
statistical properties of the phonons in the NAMR can be observed indirectly
by measuring photon correlations of the output field from the optical cavity in a similar way as in Refs.~%
\cite{DidierPRB11,RamosPRL13,CohenNat15}. In the rotating reference frame
with the frequency of the optical driving field $\omega _{a}$, the
Hamiltonian of the total quantum system is given by ($\hbar =1$)%
\begin{equation}
H_{\mathrm{T}}=H_{\mathrm{mq}}+H_{\mathrm{om}},
\end{equation}%
where%
\begin{equation}
H_{\mathrm{om}}=\Delta _{a}a^{\dag }a+ga^{\dag }a\left( b+b^{\dag }\right)
+\left( \Omega _{c}a^{\dag }+\mathrm{H.c.}\right) .
\end{equation}%
Here, $a$ and $a^{\dag }$ denote the annihilation and creation operators of
the cavity mode with the frequency $\omega _{c}$; $g$ is the vacuum optomechanical
coupling strength; $\Omega _{c}$ (assumed to be real) describes the strength
of the external driving field which satisfies the resonant condition $\Delta
_{a}\equiv \omega _{c}-\omega _{a}=\omega _{m}$. The operator for the cavity
mode can be written as the sum of its quantum fluctuation operator and
steady-state mean field as $a\rightarrow a+\alpha$, where $\alpha$ is the steady-state mean
field at the frequency $\omega_{a}$ of the driving optical field. The optical
field at the frequency of the driving optical field will be spectrally filter out and the photon correlation of the quantum fluctuation operator can be measured directly. We note that it is very challenging to measure $g_{a}^{\left( 2\right) }\left( \tau \right)$ of the quantum fluctuation by filtering out $\alpha$. However, the photon counting technique has been realized in a recent experiment~\cite{CohenNat15}. Thus, we will focus on the correlation of the quantum fluctuation operator in the following. In the strong driving
condition $|\alpha |\gg 1$, we can linearize the Hamiltonian $H_{%
\mathrm{om}}$ by only keeping the first-order terms in the small quantum
fluctuation operators, then the linearized Hamiltonian $H_{\mathrm{om}%
}^{\prime }$ under the rotating wave approximation is given as%
\begin{equation}
H_{\mathrm{om}}^{\prime }=\omega_{m}a^{\dag }a+Ga^{\dag }b+G^{\ast }ab^{\dag
},
\end{equation}%
where the effective optomechanical coupling strength $G=g\alpha$ is assumed
to be much smaller than $\omega_{m}$, i.e., $|G|\ll \omega_{m}$.

The quantum Langevin equation for the cavity mode is given by%
\begin{equation}
\frac{d}{dt}a=-\frac{\Gamma }{2}a-i\omega _{m}a-iGb+\sqrt{\Gamma }a_{\mathrm{%
vac}},
\end{equation}%
where $\Gamma $ is the damping rate of the cavity mode and the $a_{\mathrm{%
vac}}$ is the vacuum input noise with the correlation function $\langle a_{%
\mathrm{vac}}^{\dagger }(t)a_{\mathrm{vac}}(t^{\prime })\rangle =0$. We will
focus on the resolved sideband regime ($\omega _{m}\gg \Gamma $), and assume
that the damping rate $\Gamma $ of the cavity mode is much larger than the
effective optomechanical coupling strength $\left\vert G\right\vert $, the
NAMR-qubit coupling strength $J$ and the damping rates of the NAMR and
qubit, i.e.,%
\begin{equation}
\omega _{m}\gg \Gamma \gg \left\{ \left\vert G\right\vert ,J,\gamma \left(
n_{m,\mathrm{th}}+1\right) ,\kappa \left( n_{q,\mathrm{th}}+1\right)
\right\}.  \label{Eq30}
\end{equation}%
Here, $n_{m,\mathrm{th}}$ and $n_{q,\mathrm{th}}$ denote the mean thermal phonon and photon numbers of the NAMR and qubit, respectively. Then we can obtain the relation between the cavity mode and the
NAMR mode~\cite{CohenNat15,RamosPRL13,JahnePRA09,XWXuPRA15}
\begin{equation}
a=-i\frac{2G}{\Gamma }b+f_{\mathrm{vac}}
\end{equation}%
with the noise term%
\begin{equation}
f_{\mathrm{vac}}=\frac{2}{\sqrt{\Gamma }}\int_{-\infty }^{t}\left[ a_{%
\mathrm{vac}}e^{\left( \frac{\Gamma }{2}+i\omega _{m}\right) \left( \tau
-t\right) }\right] d\tau .
\end{equation}%
Since $f_{\mathrm{vac}}$ is the vacuum input, the we have
\begin{equation} \label{Eq33}
g_{b}^{\left( 2\right) }\left( \tau \right) \approx g_{a}^{\left( 2\right)
}\left( \tau \right) =\frac{\left\langle a^{\dag }\left( t\right) a^{\dag
}\left( t+\tau \right) a\left( t+\tau \right) a\left( t\right) \right\rangle
}{\left\langle a^{\dag }\left( t\right) a\left( t\right) \right\rangle ^{2}}.
\end{equation}%
by using the correlation relation $\langle a_{%
\mathrm{vac}}^{\dagger }(t)a_{\mathrm{vac}}(t^{\prime })\rangle =0$.
Equation (\ref{Eq33}) shows that phonon blockade in the NAMR can be detected by measuring the photon
correlations of the output field from the cavity mode. Note that the
effective mechanical damping rate $\widetilde{\gamma }$ and mean thermal
phonon number $\widetilde{n}_{m,\mathrm{th}}$ by adiabatically eliminating the optical cavity field under the condition [Eq.~(\ref{Eq30})] are given by~\cite%
{RamosPRL13,WilsonRaePRL07,MarquardtPRL07,LiYPRB08,XWXuPRA15}%
\begin{equation}  \label{Eq34}
\widetilde{\gamma }=\gamma +\gamma _{\mathrm{om}},
\end{equation}%
\begin{equation}  \label{Eq35}
\widetilde{n}_{m,\mathrm{th}}=\frac{\gamma n_{m,\mathrm{th}}+\gamma _{%
\mathrm{om}}n_{\mathrm{om}}}{\gamma +\gamma _{\mathrm{om}}},
\end{equation}%
where%
\begin{equation}
\gamma _{\mathrm{om}}=\frac{4\left\vert G\right\vert ^{2}}{\Gamma }\frac{%
16\omega _{m}^{2}}{\Gamma ^{2}+16\omega _{m}^{2}},
\end{equation}%
\begin{equation}
n_{\mathrm{om}}=\frac{\Gamma ^{2}}{16\omega _{m}^{2}},
\end{equation}%
are the induced mechanical damping and mean phonon number due to the
optomechanical coupling. There is an optomechanically induced
frequency shift $\delta \omega =8\left\vert G\right\vert ^{2}\omega
_{m}/(\Gamma ^{2}+16\omega _{m}^{2})$ of the mechanical resonator, which can be neglected in the
condition $\left\vert G\right\vert \ll \Gamma \ll \omega _{m}$.

\begin{figure}[tbp]
\includegraphics[bb=27 201 557 618, width=8.5 cm, clip]{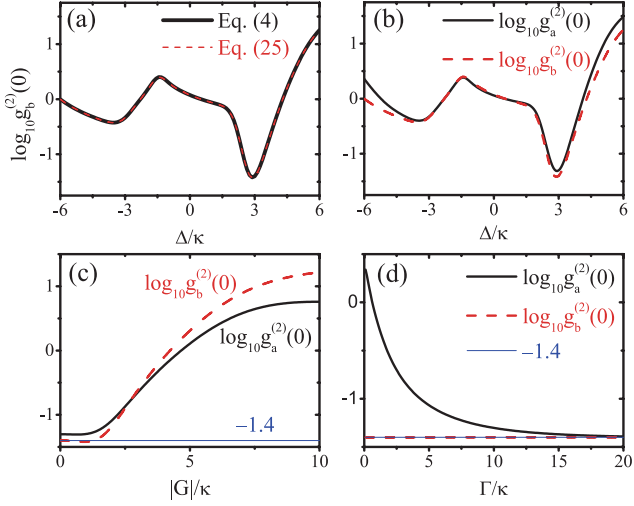}
\caption{(Color online) (a) $\log_{10}g_{b}^{\left( 2\right) }\left( 0\right)$ is
plotted as a function of the detuning $\Delta/\protect%
\kappa $ by using Eq.~(\ref{Eq6}) (black solid line) and by using Eq.~(\ref{Eq38}) (red dashed line) for $|G|=\protect\kappa/10$ and $\Gamma=10\protect\kappa$. $\log_{10}g_{a}^{\left( 2\right) }\left( 0\right)$ (black solid line) and $\log_{10}g_{b}^{\left( 2\right) }\left( 0\right)$ (red dashed line) are plotted as functions of the detuning $\Delta/\kappa $ for $|G|=\protect\kappa/10$ and $\Gamma=10\protect\kappa$ in (b), as functions of $|G|/\kappa$ for $\Gamma=10\protect\kappa$ and $\Delta=3\kappa$ in (c) and as functions of $\Gamma/\kappa $ for $|G|=\protect\kappa/10$ and $\Delta=3\kappa$ in (d). The blue thin horizontal line for $-1.4$ in (c) and (d) indicates $\log_{10}g_{b}^{\left( 2\right) }\left( 0\right)$ obtained by using Eq.~(\ref{Eq6}).
The other parameters are taken as $(\protect\eta,\protect\phi,\Delta_{\mathrm{opt}})=(%
\protect\eta_{+},\protect\phi_{+},J)$, $\protect\gamma=\protect\kappa$, $\protect\varepsilon=\protect\kappa/5$, $J=J_{\mathrm{opt}}=3%
\protect\kappa$, and $n_{\mathrm{th}}=10^{-3}$.}
\label{fig11}
\end{figure}

$g_{a}^{\left( 2\right) }\left( 0\right) $ and $g_{b}^{\left( 2\right) }\left( 0\right) $ can be obtained by solving the
master equation for the total density matrix $\rho _{\mathrm{T}}$~\cite%
{Carmichael}
\begin{eqnarray} \label{Eq38}
\frac{\partial \rho _{\mathrm{T}}}{\partial t} &=&-i\left[ H_{\mathrm{T}%
}^{\prime },\rho _{\mathrm{T}}\right] +\Gamma L[a]\rho _{\mathrm{T}}  \notag
\\
&&+\gamma \left( n_{m,\mathrm{th}}+1\right) L[b]\rho _{\mathrm{T}}+\gamma
n_{m,\mathrm{th}}L[b^{\dag }]\rho _{\mathrm{T}}  \notag \\
&&+\kappa \left( n_{q,\mathrm{th}}+1\right) L[\sigma _{\_}]\rho _{\mathrm{T}%
}+\kappa n_{q,\mathrm{th}}L[\sigma _{+}]\rho _{\mathrm{T}}
\end{eqnarray}%
with the linearized total Hamiltonian $H_{\mathrm{T}}^{\prime }$ in the rotating reference frame with
the frequency $\omega _{d}$ of the driving fields as
\begin{eqnarray}
H_{\mathrm{T}}^{\prime } &=&\Delta a^{\dag }a+\Delta b^{\dag }b+\Delta
\sigma _{+}\sigma _{-}  \notag \\
&&+Ga^{\dag }b+G^{\ast }ab^{\dag }+J\left( \sigma _{+}b+b^{\dag }\sigma
_{-}\right)   \notag \\
&&+\left( \varepsilon b^{\dag }+\Omega e^{-i\phi }\sigma _{+}+\mathrm{H.c.}%
\right) ,
\end{eqnarray}%
where $L[a]\rho_{\rm T}$ is the Lindbland term for the optical cavity mode.

In Fig.~\ref{fig11}(a), we compare the results of $\log_{10}g_{b}^{\left( 2\right) }\left( 0\right)$ calculated by using Eq.~(\ref{Eq6}) (black solid line) and by using Eq.~(\ref{Eq38}) (red dashed line). It clearly shows that they agree well with each other by the parameters $\omega _{m}\gg \Gamma $, $|G|=\protect\kappa/10$ and $\Gamma=10\protect\kappa$. That is because the effect of the optical field is so small ($\gamma_{\rm om}\ll \gamma$ and $\gamma_{\rm om}n_{\rm om}\ll \gamma n_{m,\mathrm{th}}$) that we have $\widetilde{\gamma }\approx \gamma$ and $\widetilde{n}_{m,\mathrm{th}}\approx n_{m,\mathrm{th}}$ as given in Eqs.~(\ref{Eq34}) and (\ref{Eq35}).
The correlations of the photons and phonons are shown in Fig.~\ref{fig11}(b), which shows $g_{b}^{\left( 2\right) }\left( 0\right) \approx g_{a}^{\left(2\right) }\left( 0\right) $ with the same parameters as in Fig.~\ref{fig11}(a). So phonon blockade in the NAMR mode can be
observed indirectly by measuring photon correlations of the output field
from the optical cavity. Moreover, the correlation functions are plotted as functions of $|G|/\kappa$ in Fig.~\ref{fig11}(c) and as functions of $\Gamma/\kappa $ in \ref{fig11}(d). We find that when the condition in Eq.~(\ref{Eq30}) is not satisfied (e.g., $\Gamma \sim |G| \gg \kappa$ or $\Gamma \sim |G| \approx \kappa$), $\log_{10}g_{a}^{\left( 2\right) }\left( 0\right)$ (black solid line) and $\log_{10}g_{b}^{\left( 2\right) }\left( 0\right)$ calculated by using Eq.~(\ref{Eq38}) (red dashed line) and Eq.~(\ref{Eq6}) (i.e., $-1.4$ indicated by blue thin horizontal line) become quite different. These differences can be understood by noting that the influence of the optical field gradually emerges with the increasing of the effective optomechanical coupling strength $|G|$ and the photons in the cavity mode can evolve adiabatically with the phonons in the mechanical mode only when $\Gamma  \gg |G| \sim \kappa$.

\section{Conclusions}

In summary, we have studied phonon blockade in a NAMR which is
resonantly coupled to a qubit. We have shown that phonon blockade can be
induced not only by strong nonlinear interactions, corresponding to large
coupling strengths between the NAMR and the qubit (called as CPNB), but also
by the destructive interference between different paths for two-phonon
excitation with a moderate coupling strength between the NAMR and the qubit
(called as UCPNB).

Although UCPNB can appear in a moderate (even weak) NAMR-qubit coupling
regime and the phonon blockade is almost perfect, we find that the mean
phonon number is very small and it is very fragile in the thermal reservoir.
In contrast to UCPNB, CPNB can be observed with much larger mean phonon
number and more robust against the thermal noise, but it needs very strong
NAMR-qubit coupling to blockade subsequent phonons.

We also show that a strong phonon antibunching, with a moderate NAMR-qubit coupling, large mean
phonon number and robust against the thermal noise, can be achieved by applying
two external driving fields to the NAMR and qubit, respectively. The phonon
blockade in the NAMR can be observed indirectly by measuring photon
correlations of the output field from the optical cavity which is optomechanically
coupled to the NAMR. Our proposal provide a way to observe phonon blockade in the NAMR via its resonant coupling with a qubit.

Finally, let us discuss the experimental feasibility of our proposal.
In a recent experiment for a superconducting phase qubit coupled to a NAMR by a Jaynes-Cummings model~\cite{OConnellNat10}, the NAMR-qubit coupling strength $J/2\pi$ is about $124$~MHz, the frequency of the NAMR $\omega_{m}/2\pi$ is about $6$~GHz, the qubit frequency $\omega_{0}/2\pi$ can be set between $5$ and $10$~GHz, the damping rate of the NAMR $\gamma/2\pi$ is about $26$~MHz, the damping rate of the phase qubit $\kappa/2\pi$ is about $9$~MHz, and the environmental temperature can be cooled to $25$~mK by a dilution refrigerator.
All the parameters used in our paper is within the reach of this experiment~\cite{OConnellNat10}.
Moreover, in another recent experiment~\cite{CohenNat15}, the correlations of phonons for a NAMR in an optomechanical system has been measured by detecting the correlations of the emitted photons from the optical cavity, where the vacuum optomechanical coupling strength $g/2\pi$ is $645$~kHz and the damping rate of the optical cavity $\Gamma/2\pi$ is $817$~MHz, which satisfy the condition in Eq.~(\ref{Eq30}) for phonon correlation measuring. Thus, if the superconducting NAMR-qubit system can be combined with the optomechanical system as an electro-optomechanical system~\cite{BochmannNPy13,BagciNat14,AndrewsNPy14}, then the phonon blockade in the NAMR should be observed in our proposed system.

\vskip 2pc \leftline{\bf Acknowledgement}

X.W.X. thank Yong Li, Hui Wang and Yan-Jun Zhao for fruitful discussions.
X.W.X. is supported by the
National Natural Science Foundation of China (NSFC) under Grants No.11604096
and the Startup Foundation for Doctors of East China Jiaotong University
under Grant No. 26541059. A.X.C. is supported by NSFC under Grant No.
11365009. Y.X.L. is supported by the National Basic Research Program of China(973 Program) under
Grant No. 2014CB921401, the Tsinghua University Initiative
Scientific Research Program, and the Tsinghua National Laboratory for
Information Science and Technology (TNList) Cross-discipline Foundation.

\appendix

\section{Derivation of optimal equation for UCPNB}\label{OpCo}

To obtain the optimal conditions for UCPNB in the limit $T\rightarrow 0$, following the method given in
Ref.~\cite{BambaPRA11}, we expand the wave function with the Ansatz~%
\begin{eqnarray}
\left\vert \psi \right\rangle &=&C_{0,g}\left\vert 0,g\right\rangle
+C_{0,e}\left\vert 0,e\right\rangle +C_{1,g}\left\vert 1,g\right\rangle
\notag  \label{Eq7} \\
&&+C_{1,e}\left\vert 1,e\right\rangle +C_{2,g}\left\vert 2,g\right\rangle
+\cdots .
\end{eqnarray}%
Here, $\left\vert n,g\right\rangle $ ($\left\vert n,e\right\rangle $)
represents that there are $n$ phonons in the mechanical mode and the qubit
in the ground (excited) state. The coefficients $|C_{n,g}|^2$ and $|C_{n,e}|^2$ denote occupying probabilities in states $\left\vert n,g\right\rangle $ and $\left\vert n,e\right\rangle $ respectively.
Under weak pumping conditions ($C_{0,g}\approx 1\gg \left\{
C_{0,e},C_{1,g}\right\} \gg \left\{ C_{1,e},C_{2,g}\right\} \gg \cdots $),
the optimal condition for UCPNB can be obtained for $C_{2,g}=0$.

Substituting the wave function in Eq.~(\ref%
{Eq7}) and the Hamiltonian in Eq.~(\ref{Eq2}) into the Schr\"{o}dinger
equation, by taking account of the dampings of the qubit and mechanical mode,
we can obtain the dynamical equations for the amplitudes $C_{n,g}$\ ($%
C_{n,e}$).
The steady-state solutions for amplitudes $C_{1,g}$ and $C_{0,e}$ are
determined by%
\begin{eqnarray}
0 &=&\left( \Delta -i\frac{\kappa }{2}\right) C_{0,e}+JC_{1,g}+\Omega
e^{-i\phi }C_{0,g},  \label{Eq8} \\
0 &=&\left( \Delta -i\frac{\gamma }{2}\right) C_{1,g}+JC_{0,e}+\varepsilon
C_{0,g},  \label{Eq9}
\end{eqnarray}%
and the steady-state solutions for amplitudes $C_{2,g}$ and $C_{1,e}$ are
determined by%
\begin{eqnarray}
0 &=&\left( 2\Delta -i\frac{\kappa +\gamma }{2}\right) C_{1,e}+\sqrt{2}%
JC_{2,g}  \notag  \label{Eq10} \\
&&+\varepsilon C_{0,e}+\Omega e^{-i\phi }C_{1,g}, \\
0 &=&\left( 2\Delta -i\gamma \right) C_{2,g}+\sqrt{2}JC_{1,e}+\sqrt{2}%
\varepsilon C_{1,g}.  \label{Eq11}
\end{eqnarray}%
From Eqs.~(\ref{Eq8}) and (\ref{Eq9}), we obtain
\begin{equation}
C_{0,e}=\frac{\varepsilon J-\Omega \left( \Delta -i\frac{\gamma }{2}\right)
e^{-i\phi }}{\left( \Delta -i\frac{\kappa }{2}\right) \left( \Delta -i\frac{%
\gamma }{2}\right) -J^{2}}C_{0,g},  \label{Eq12}
\end{equation}%
\begin{equation}
C_{1,g}=\frac{J\Omega e^{-i\phi }-\varepsilon \left( \Delta -i\frac{\kappa }{%
2}\right) }{\left( \Delta -i\frac{\kappa }{2}\right) \left( \Delta -i\frac{%
\gamma }{2}\right) -J^{2}}C_{0,g}.  \label{Eq13}
\end{equation}%
Optimal condition for phonon blockade (i.e., $g_{b}^{\left( 2\right)
}\left( 0\right) \rightarrow 0$) can be derived by substituting Eqs.~(\ref%
{Eq12}) and (\ref{Eq13}) into Eqs.~(\ref{Eq10}) and (\ref{Eq11}) with $%
C_{2,g}=0$, and then we obtain the linear equations for $C_{1,e}$ and $%
C_{0,g}$%
\begin{equation}
\left(
\begin{array}{cc}
M_{11} & M_{12} \\
M_{21} & M_{22}%
\end{array}%
\right) \left(
\begin{array}{c}
C_{1,e} \\
C_{0,g}%
\end{array}%
\right) =0,
\end{equation}%
with
\begin{eqnarray}
M_{11} &=&2\Delta _{\mathrm{opt}}-i\frac{\kappa +\gamma }{2}, \\
M_{12} &=&\frac{J_{\mathrm{opt}}}{D}\left( \varepsilon ^{2}+\Omega
^{2}e^{-i2\phi }\right)  \notag \\
&&-\frac{1}{D}\left( 2\Delta _{\mathrm{opt}}-i\frac{\kappa +\gamma }{2}%
\right) \varepsilon \Omega e^{-i\phi }, \\
M_{21} &=&\sqrt{2}J_{\mathrm{opt}}, \\
M_{22} &=&\frac{\sqrt{2}}{D}J_{\mathrm{opt}}\varepsilon \Omega e^{-i\phi }-%
\frac{\sqrt{2}}{D}\left( \Delta _{\mathrm{opt}}-i\frac{\kappa }{2}\right)
\varepsilon ^{2}, \\
D &=&\left( \Delta _{\mathrm{opt}}-i\frac{\kappa }{2}\right) \left( \Delta _{%
\mathrm{opt}}-i\frac{\gamma }{2}\right) -J_{\mathrm{opt}}^{2},
\end{eqnarray}%
where $\Delta _{\mathrm{opt}}$ and $J_{\mathrm{opt}}$ are the optimal
parameters for phonon blockade (i.e., $g_{b}^{\left( 2\right) }\left(
0\right) \rightarrow 0$) with the corresponding parameters of the external
driving fields ($\varepsilon $, $\Omega $ and $\phi $). The condition for $%
C_{1,e}$ and $C_{0,g}$ with nontrivial solutions is that the determinant of
the coefficient matrix equals zero, that is 
\begin{equation}
M_{11}M_{22}-M_{12}M_{21}=0.
\end{equation}
We derive the equation for optimal phonon antibunching given in Eq.~(\ref{Eq19}) by simplifying the above equation and obtain a second-order equation in the strength ratio $\eta \equiv\Omega /\varepsilon $ with coefficients $A_{j}$ ($j=1,2,3$) defined in Eqs.~(\ref{Eqa6})-(\ref{Eqa8}).

\bibliographystyle{apsrev}
\bibliography{ref}

\end{document}